\documentclass{article}
\usepackage{spconf,amsmath}
\usepackage{cite}
\usepackage{setspace}
\usepackage{bm}
\usepackage{algorithmic}
\usepackage{array}
\usepackage{amssymb}
\usepackage{graphicx}
\usepackage{url}
\usepackage{comment}
\usepackage{fancyhdr}

\title{Blood Oxygen Saturation Estimation from Facial Video\\ via DC and AC components of Spatio-temporal Map}
\name{Yusuke Akamatsu$^{\dagger}$ \qquad Yoshifumi Onishi \qquad Hitoshi Imaoka}
\address{Biometrics Research Laboratories, NEC Corporation, Japan}

\begin{document}
\ninept
\maketitle
\setlength{\abovedisplayskip}{2mm}
\setlength{\belowdisplayskip}{2mm}
%
\begin{abstract}
Peripheral blood oxygen saturation (SpO2), an indicator of oxygen levels in the blood, is one of the most important physiological parameters.
Although SpO2 is usually measured using a pulse oximeter, non-contact SpO2 estimation methods from facial or hand videos have been attracting attention in recent years.
In this paper, we propose an SpO2 estimation method from facial videos based on convolutional neural networks (CNN).
Our method constructs CNN models that consider the direct current (DC) and alternating current (AC) components extracted from the RGB signals of facial videos, which are important in the principle of SpO2 estimation.
Specifically, we extract the DC and AC components from the spatio-temporal map using filtering processes and train CNN models to predict SpO2 from these components.
We also propose an end-to-end model that predicts SpO2 directly from the spatio-temporal map by extracting the DC and AC components via convolutional layers.
Experiments using facial videos and SpO2 data from 50 subjects demonstrate that the proposed method achieves a better estimation performance than current state-of-the-art SpO2 estimation methods.
\end{abstract}

\renewcommand{\thefootnote}{\fnsymbol{footnote}}
\footnote[0]{$^{\dagger}$Contact author : yusuke-akamatsu@nec.com}
\renewcommand{\thefootnote}{\arabic{footnote}}

\section{Introduction}
\label{sec:intro}
Peripheral blood oxygen saturation (SpO2) is the ratio of oxygenated hemoglobin (HbO2) with respect to total hemoglobin, indicating the levels of oxygen supply in the blood.
Oxygen saturation is an essential physiological parameter in the treatment of chronic cardiovascular and respiratory diseases.
Especially in patients infected with COVID-19, it has been reported that hypoxemia can occur before breathing difficulties are observed~\cite{couzin2020mystery,starr2020pulse}.
Therefore, continuous SpO2 monitoring is becoming increasingly important. 
SpO2 is usually measured using a pulse oximeter attached to the fingertip.
However, contact-based pulse oximeters can cause discomfort, increase the risk of infection, and are not suitable for continuous monitoring.
In addition, they have not been widely used in households.
Recently, non-contact SpO2 estimation methods using facial or hand videos have been proposed~\cite{humphreys2006cmos,kong2013non,shao2015noncontact,tarassenko2014non,guazzi2015non,rahman2019non,casalino2020mhealth,sun2021robust,wei2021analysis,DBLP:journals/corr/abs-2107-05087,tian2022multi,akamatsu2022heart}, which may enable measurement using a tablet or smartphone.
These video-based methods enable remote monitoring of SpO2 in video meetings and telemedicine scenarios.

Video-based SpO2 estimation methods can be divided on the basis of the type of camera they use: either special cameras that can capture specific wavelength bands~\cite{humphreys2006cmos,kong2013non,shao2015noncontact} or RGB cameras such as webcams and smartphone cameras~\cite{tarassenko2014non,guazzi2015non,rahman2019non,casalino2020mhealth,sun2021robust,wei2021analysis,DBLP:journals/corr/abs-2107-05087,tian2022multi,akamatsu2022heart}.
Special cameras can capture wavelengths suitable for SpO2 estimation, but such cameras are not widely used.
Hence, an RGB camera-based SpO2 estimation method is desired.
Previous RGB camera-based methods~\cite{tarassenko2014non,guazzi2015non,rahman2019non,casalino2020mhealth} estimated SpO2 from facial videos based on the ratio-of-ratios (RoR)~\cite{webster1997design}, which is similar to the estimation principle used in pulse oximeters.
RoR-based methods~\cite{tarassenko2014non,guazzi2015non,rahman2019non,casalino2020mhealth} calculate the ratio between direct current (DC) and alternating current (AC) components of red and blue channel time-series data acquired from videos (see Section~\ref{sec:pre}).
Several methods have extended RoR by incorporating mechanisms for stably extracting the DC and AC components of the time-series data~\cite{wei2021analysis} and regression models using three channels (red, green, and blue)~\cite{sun2021robust,tian2022multi}.
Furthermore, several authors have constructed convolutional neural networks (CNN) to predict SpO2 from three-channel time-series data and estimated SpO2 from facial~\cite{akamatsu2022heart} or hand~\cite{DBLP:journals/corr/abs-2107-05087} videos.
CNN can extract effective features for SpO2 estimation from time-series data and has outperformed the RoR-based methods.
However, CNN-based methods~\cite{DBLP:journals/corr/abs-2107-05087,akamatsu2022heart} are not designed to explicitly extract the DC and AC components, which are important in the principle of SpO2 estimation.
Therefore, building CNN models that extract the DC and AC components may improve the SpO2 estimation performance.

\begin{figure*}[t]
\begin{center}
\includegraphics[scale=0.4]{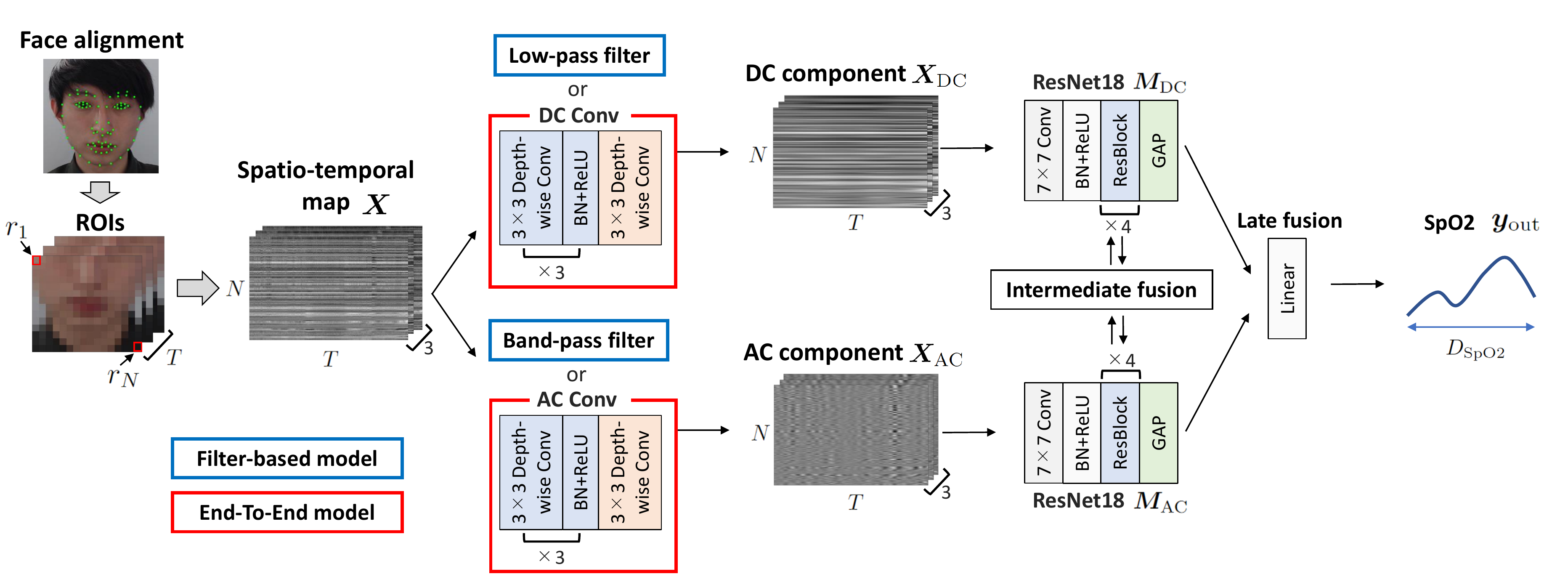}
\end{center}
\vspace{-15pt}
\caption{Overview of proposed method. First, we construct a spatio-temporal map and extract DC and AC components using filtering processes (low-pass filter and band-pass filter) or convolutional layers (DC Conv and AC Conv). The extracted DC and AC components are then input to two ResNet18 networks that are fused via intermediate and late fusion to output SpO2.}
\label{fig:overview}
\end{figure*}

In this paper, we propose a CNN-based method that considers the DC and AC components of three-channel time-series data to estimate SpO2 from facial videos.
First, we construct a spatio-temporal map using three-channel time-series data in multiple regions of interest (ROIs) obtained from facial videos.
Next, we extract the DC and AC components from the spatio-temporal map by filtering processes ({\it i.e.}, low-pass and band-pass filters).
The DC and AC components are then respectively input to two CNN models that are fused via intermediate and late fusion to output SpO2.
We also propose an end-to-end model that outputs SpO2 directly from the spatio-temporal map by training convolutional layers to extract the DC and AC components without filtering. 
The end-to-end model trains the DC and AC components extraction and SpO2 estimation simultaneously, which enables the effective features for SpO2 estimation to be automatically acquired from the spatio-temporal map.
We conducted experiments with data from 50 subjects and confirmed that the proposed methods outperform the RoR, regression, and CNN-based methods that do not consider the DC and AC components.

\section{Preliminary: ratio-of-ratios}
\label{sec:pre}
SpO2 represents the oxygen saturation in peripheral capillaries and is calculated by
\begin{equation}
    \rm{SpO2} = \frac{\rm{HbO2}}{\rm{HbO2}+\rm{Hb}}\times 100\%,
\end{equation}
where $\rm{HbO2}$ is oxygenated hemoglobin and $\rm{Hb}$ is deoxygenated hemoglobin~\cite{kong2013non}.
The pulse oximeter measures SpO2 based on the difference in absorbance by HbO2 and Hb at two wavelengths: $\lambda_1$ = 660 nm (red) and $\lambda_2$ = 940 nm (infrared).
Specifically, the absorption coefficients of Hb and HbO2 are greatly different at $\lambda_1$, whereas their absorption coefficients are approximately equivalent at $\lambda_2$~\cite{shao2015noncontact,DBLP:journals/corr/abs-2107-05087}.
In general, the light reflected from skin can be divided into pulsatile and non-pulsatile components, which respectively correspond to the AC and DC components~\cite{wei2021analysis}.
By normalizing the AC component by the DC component at $\lambda_1$ and $\lambda_2$, the pulsatile absorbance rates $R_{\lambda_1}$ and $R_{\lambda_2}$ are obtained as
\begin{equation}
    R_{\lambda_1} = \frac{\rm{AC}_{\lambda_1}}{\rm{DC}_{\lambda_1}},\quad R_{\lambda_2} = \frac{\rm{AC}_{\lambda_2}}{\rm{DC}_{\lambda_2}}.
\end{equation}
The ratio of absorbance at two wavelengths is defined as the ratio-of-ratios (RoR), and RoR can be considered as nearly linear with respect to SpO2~\cite{shao2015noncontact,tian2022multi}
\begin{align}
    &{\rm SpO2} \approx A \cdot {\rm RoR} + B, \nonumber \\
    &{\rm RoR} = \frac{R_{\lambda_1}}{R_{\lambda_2}} = \frac{\rm{AC}_{\lambda_1}/\rm{DC}_{\lambda_1}}{\rm{AC}_{\lambda_2}/\rm{DC}_{\lambda_2}},
\end{align}
where $A$ and $B$ are linear parameters.
Previous SpO2 estimation methods using RGB cameras~\cite{tarassenko2014non,guazzi2015non,rahman2019non,casalino2020mhealth} utilized the blue channel as $\lambda_2$ instead of the infrared in pulse oximeters.
This is because the absorption coefficients of Hb and HbO2 are approximately equivalent in the blue channel, as well as in the infrared.

\section{Method: CNN considering DC and AC components}
The RoR in Eq. (3) can be described as
\begin{align}
    {\rm RoR} =  \frac{\rm{AC}_{\lambda_1}}{\rm{AC}_{\lambda_2}} \cdot \frac{\rm{DC}_{\lambda_2}}{\rm{DC}_{\lambda_1}} =
    R_{\rm AC} \cdot R_{\rm DC}.
\end{align}
We construct two CNN models corresponding to $R_{\rm AC}$ and $R_{\rm DC}$, respectively, and fuse them to predict SpO2.
This construction improves the SpO2 estimation performance over the previous CNN models~\cite{DBLP:journals/corr/abs-2107-05087,akamatsu2022heart} that do not consider the DC and AC components.
The RoR~\cite{tarassenko2014non,guazzi2015non,rahman2019non,casalino2020mhealth} uses the mean value of the time-series data of red and blue channels over a certain time period ({\it e.g.}, 10 seconds~\cite{tarassenko2014non}) as the DC component.
It also utilizes the standard deviation~\cite{rahman2019non,casalino2020mhealth} or mean peak-to-trough heights~\cite{tarassenko2014non} of the time-series data as the AC component.
Thus, RoR does not consider the temporal information of the DC and AC components and the corresponding SpO2 in a certain time period.
In contrast, the proposed method can take temporal information into account by using a spatio-temporal map.
Figure~\ref{fig:overview} shows an overview of the proposed method.

\subsection{DC and AC Components of Spatio-Temporal Map}
First, we apply a face alignment method~\cite{imaoka2021future} to the facial videos and detect the facial feature points.
We then define $16 \times 14$ blocks within the rectangle from below the eyes and create multiple ROIs by arranging them.
ROI including the region below the eyes is also used in~\cite{kong2013non}.
Next, a spatio-temporal map $\bm{X} \in \mathbb{R}^{3 \times N \times T}$ is constructed using the time-series data of three channels (red, green, and blue) in multiple ROIs.
Here, $3$, $N$, and $T$ are the number of channels, ROIs including $r_1, \cdots, r_N$, and frames of the video, respectively ($N = 224$ and $T = 300$ in our case). 
Inspired by previous studies~\cite{sun2021robust,tian2022multi,DBLP:journals/corr/abs-2107-05087,akamatsu2022heart}, we use three channels instead of just the two channels (red and blue) used in RoR.
In addition, while a single ROI is used in~\cite{sun2021robust,tian2022multi,DBLP:journals/corr/abs-2107-05087}, the proposed method uses multiple ROIs.
By extracting RGB signals from multiple ROIs on the face, we can construct a more robust model compared to those using a single ROI.

For the time-series data $\bm{x}_{i,n}$ of the $n$th ROI of channel $i$, our filter-based model extracts the DC component and AC component by using a low-pass filter (below 0.3 Hz) and a bandpass filter (0.75 -- 2.5 Hz), respectively.
By applying these filters to all channels and ROIs, we extract DC and AC components of spatio-temporal maps $\bm{X}_{\rm DC} \in \mathbb{R}^{3 \times N \times T}$ and $\bm{X}_{\rm AC} \in \mathbb{R}^{3 \times N \times T}$.
We also propose an end-to-end model that extracts the DC and AC components from the spatio-temporal map by using convolutional layers (DC and AC Conv in Fig.~\ref{fig:overview}).
The DC and AC Conv consist of depth-wise convolutions, batch normalization (BN), and ReLU activation~\cite{nair2010rectified}.
By using depth-wise convolutions that have independent convolutions for each channel, we can calculate the DC and AC components per channel as shown in Eq. (4).

\subsection{Fusion of CNN Models and Loss Functions}
We input DC and AC components of spatio-temporal maps $\bm{X}_{\rm DC}$ and $\bm{X}_{\rm AC}$ into CNN models $\bm{M}_{\rm DC}$ and $\bm{M}_{\rm AC}$, respectively.
We use ResNet18~\cite{he2016deep} pre-trained on ImageNet~\cite{deng2009imagenet} as $\bm{M}_{\rm DC}$ and $\bm{M}_{\rm AC}$.
ResNet18 is a commonly utilized CNN network in various computer vision tasks and is also used for heart rate estimation from facial videos~\cite{niu2018synrhythm,niu2019rhythmnet}.
Since the number of samples including facial videos and SpO2 is limited, we improve the generalization performance by using a model that was pre-trained on a large-scale ImageNet.
Next, we fuse $\bm{M}_{\rm DC}$ and $\bm{M}_{\rm AC}$ via intermediate and late fusion to output SpO2.
In the proposed method, we utilize a multi-modal transfer module (MMTM)~\cite{joze2020mmtm} as the intermediate fusion.
MMTM fuses the intermediate layers of arbitrary CNN models with a squeeze and excitation (SE) module~\cite{hu2018squeeze}, enabling information exchange between the models.
After intermediate fusion in the four main blocks of ResNet18 ({\it i.e.}, ResBlock), the outputs of global average pooling (GAP) are fused via late fusion.
With not only late fusion but also intermediate fusion, we can combine the relevant information from three channels in the AC and DC components at an intermediate level, as in Eq. (4).

After late fusion, the proposed method outputs SpO2 values $\bm{y}_{\rm out} \in \mathbb{R}^{{D}_{\rm SpO2}}$ via a linear layer.
Here, ${D}_{\rm SpO2}$ is the dimension of the output vector and is the number of seconds of predicted SpO2 values (${D}_{\rm SpO2} = 10$ in our case).
If we suppose that $\bm{y}_{\rm GT} \in \mathbb{R}^{{D}_{\rm SpO2}}$ is the SpO2 values of ground truth, our filter-based model is trained with the following loss function:
\begin{equation}
    \mathcal{L}_{\rm SpO2} = {\rm MSE}(\bm{y}_{\rm out},\bm{y}_{\rm GT}) + {\rm NegCorr}(\bm{y}_{\rm out},\bm{y}_{\rm GT}),
\end{equation}
where MSE denotes mean squared error and NegCorr denotes negative Pearson correlation coefficient.
While MSE minimizes the error of the estimated SpO2, NegCorr maximizes its trend similarity.
In addition, our end-to-end model uses the errors between the DC component $\hat{\bm{X}}_{\rm DC}$ and AC component $\hat{\bm{X}}_{\rm AC}$ estimated by DC and AC Conv and $\bm{X}_{\rm DC}$ and $\bm{X}_{\rm AC}$ extracted by the filtering processes.
Note that while we need to extract $\bm{X}_{\rm DC}$ and $\bm{X}_{\rm AC}$ by the filtering processes during model training, the extraction is not required during inference.
The loss function of the end-to-end model is described as
\begin{align}
    \mathcal{L}_{\rm EndToEnd} = \mathcal{L}_{\rm SpO2} +  \alpha \cdot \{{\rm MSE}&(\hat{\bm{X}}_{\rm DC},\bm{X}_{\rm DC})+ \nonumber \\
    &{\rm MSE}(\hat{\bm{X}}_{\rm AC},\bm{X}_{\rm AC})\},
\end{align}
where $\alpha$ is a hyperparameter that controls the loss function of the DC and AC components estimation.
By using $\alpha$, we can control the strength of the SpO2 estimation and the DC and AC components estimation.

\begin{figure*}[t]
\begin{center}
\includegraphics[scale=0.34]{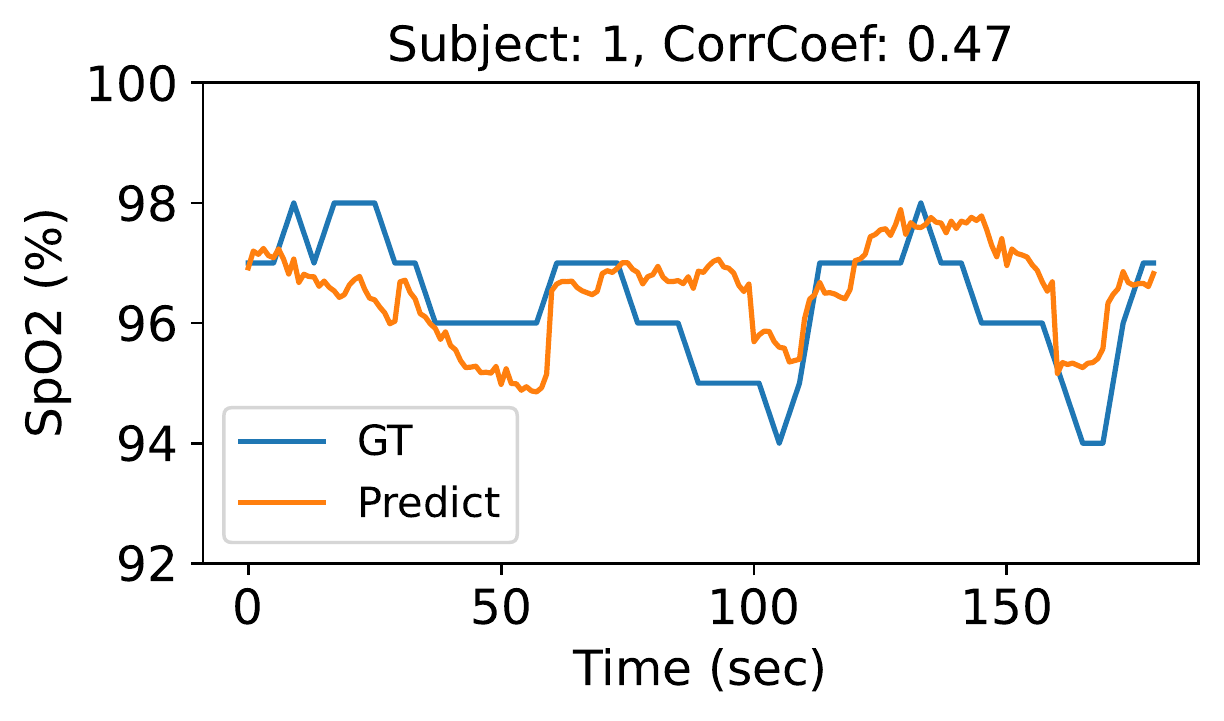}
\includegraphics[scale=0.34]{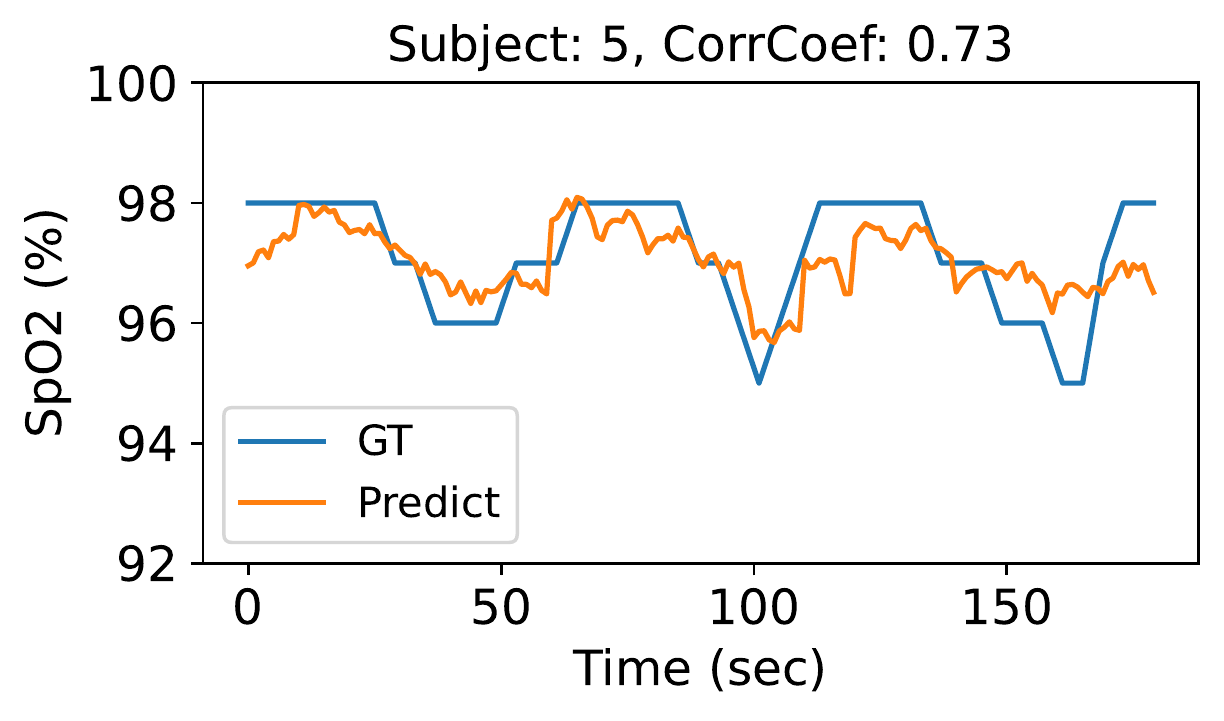}
\includegraphics[scale=0.34]{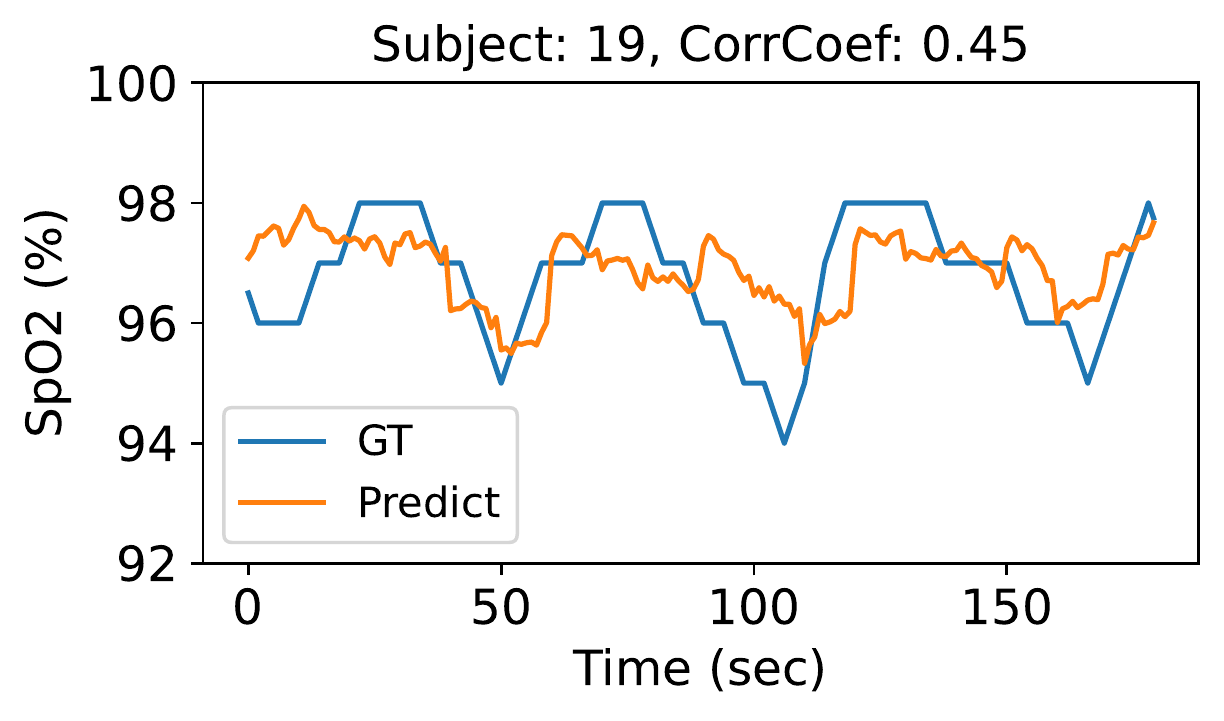}
\includegraphics[scale=0.34]{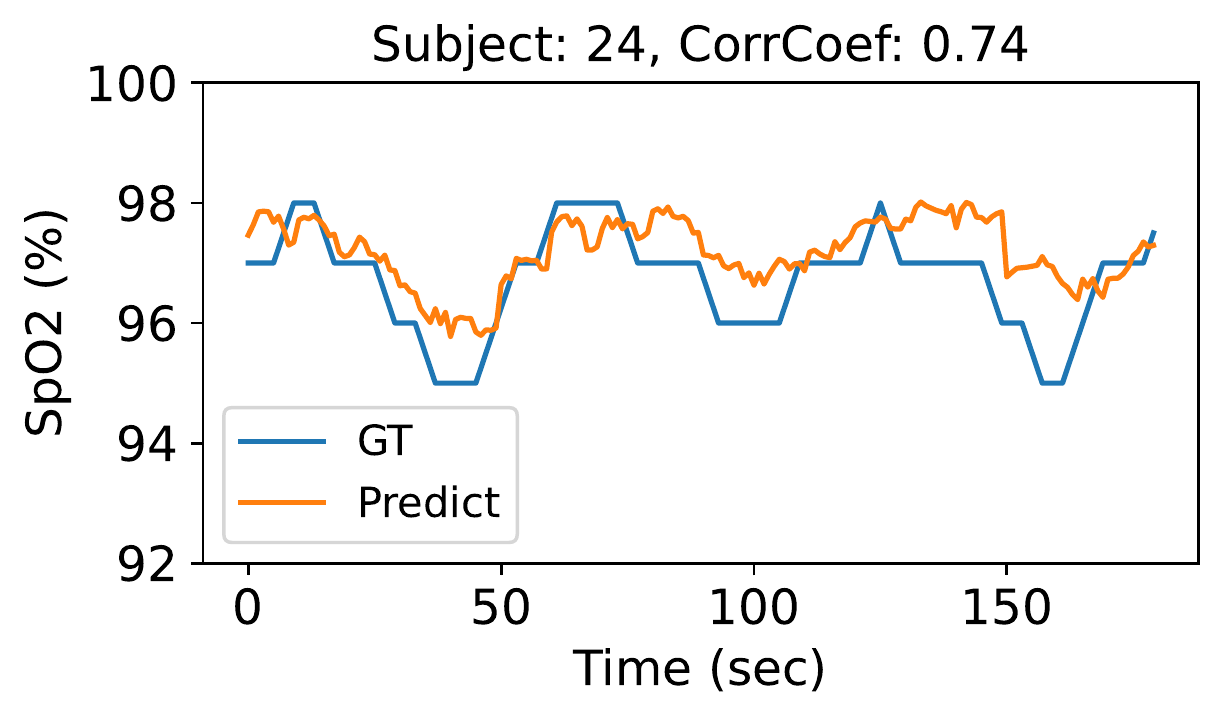}

\includegraphics[scale=0.34]{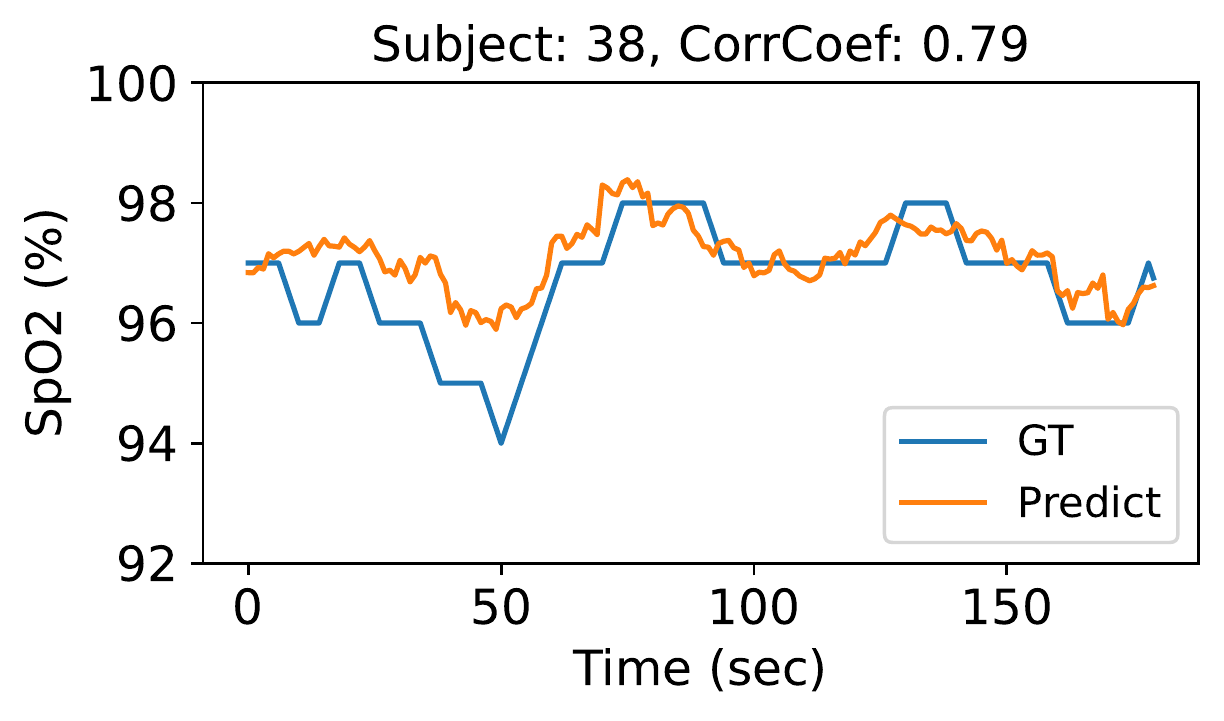}
\includegraphics[scale=0.34]{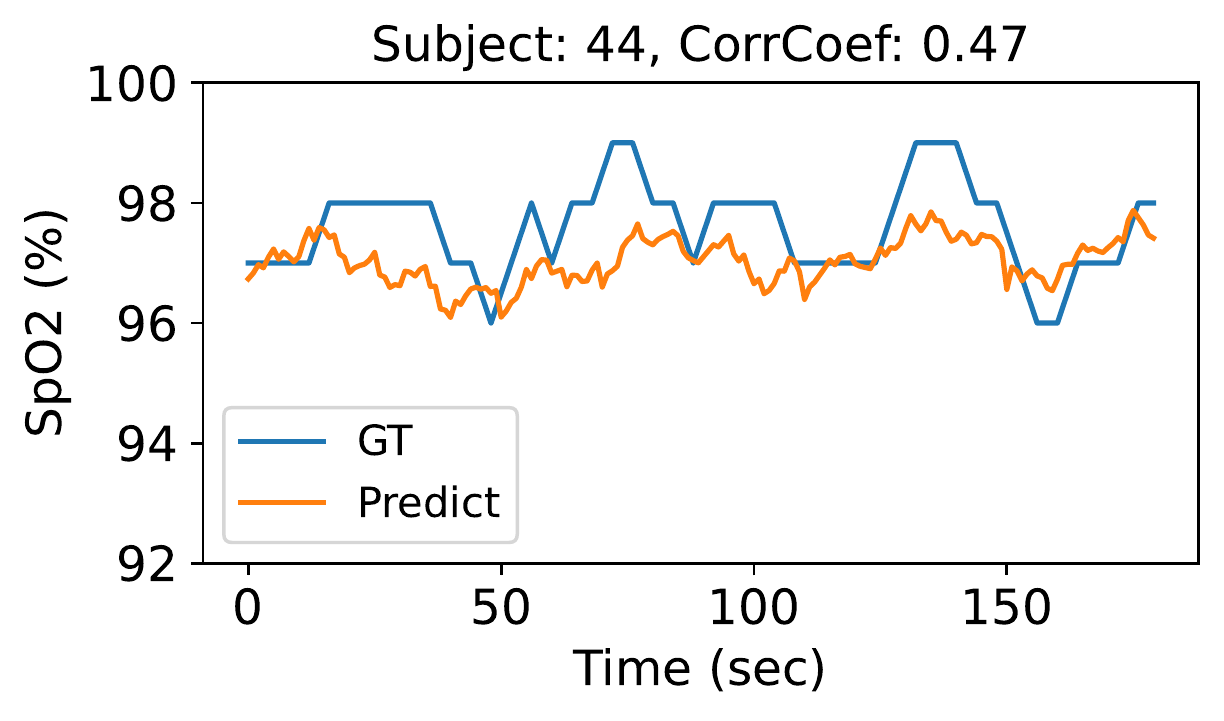}
\includegraphics[scale=0.34]{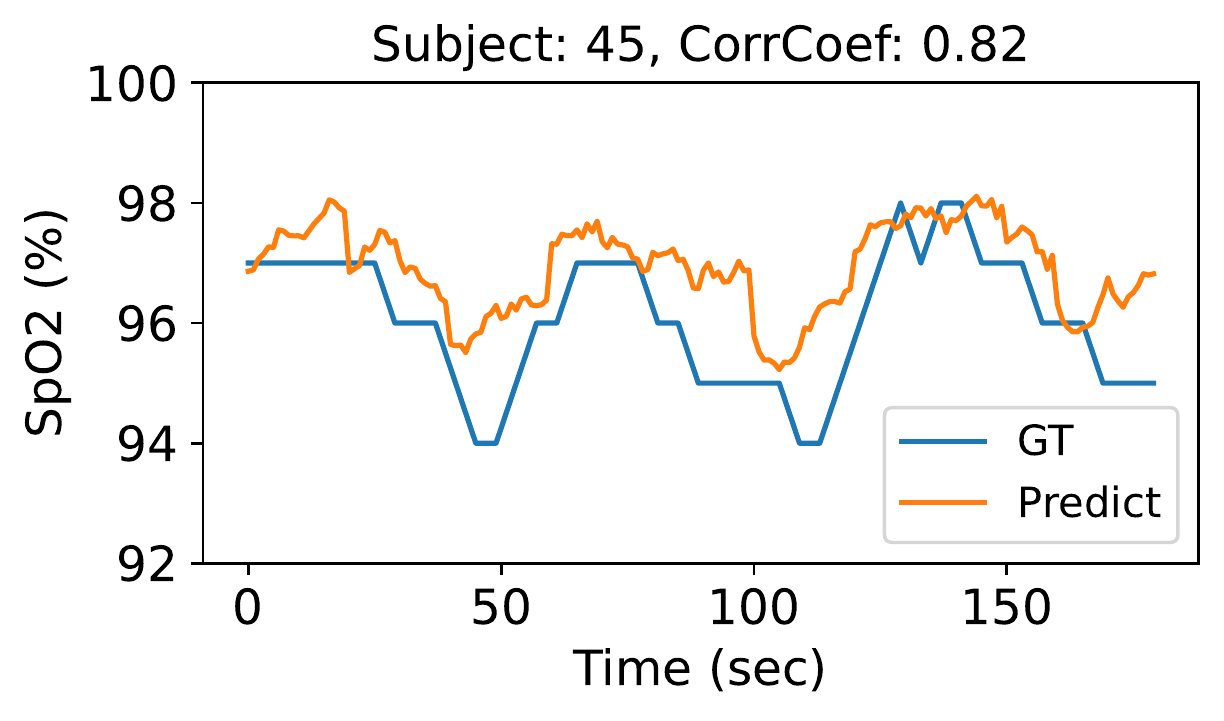}
\includegraphics[scale=0.34]{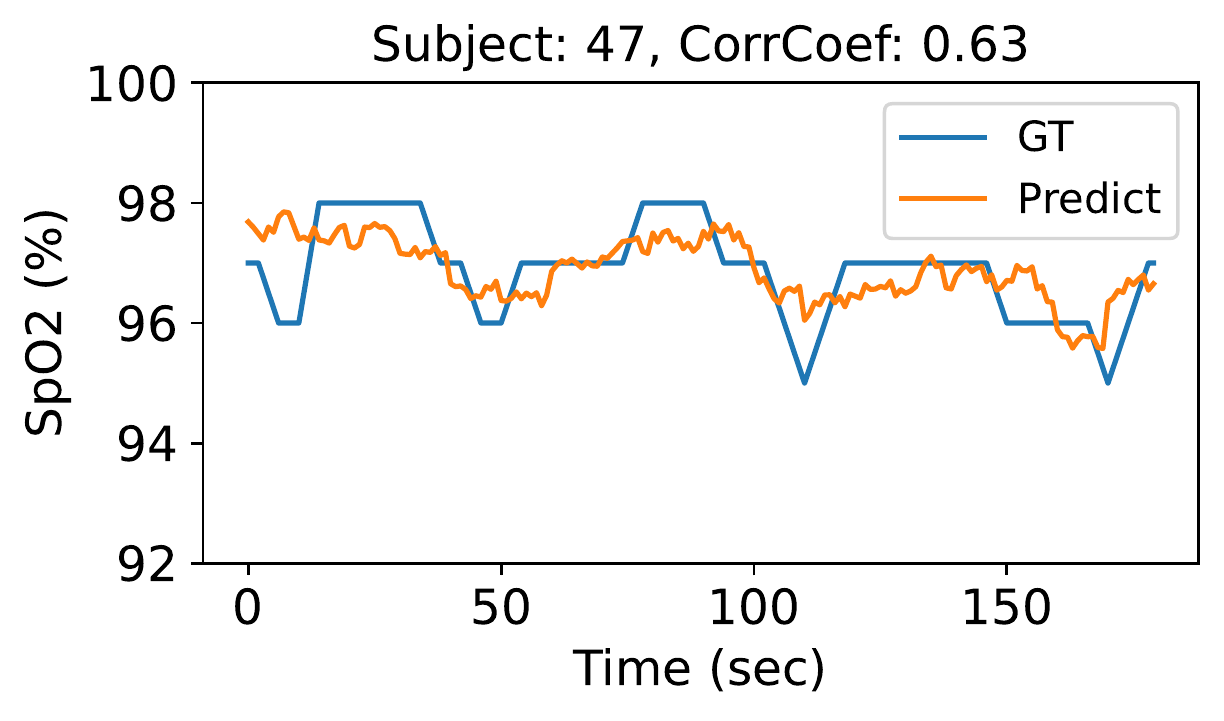}

\end{center}
\vspace{-20pt}
\caption{Predicted SpO2 and ground truth data for several subjects using CNN-EndToEnd. The horizontal axis indicates time (seconds) and the vertical axis indicates SpO2 value (\%).}
\label{fig:spo2}
\vspace{-10pt}
\end{figure*}

\section{Experiments}
\subsection{Dataset and Experimental Setup}
The proposed method was evaluated on a self-collected dataset consisting of facial videos and SpO2 data from 50 subjects (ages from 20s to 50s).
We obtained three minutes of facial videos and corresponding SpO2 data for each subject.
As in previous studies~\cite{shao2015noncontact,wei2021analysis,DBLP:journals/corr/abs-2107-05087}, subjects were asked to hold their breath quietly in order to obtain low SpO2 data.
One cycle consisted of approximately 30 seconds of breath holding and 30 seconds of rest (60 seconds in total), and the cycles were repeated three times within a three-minute period.
If subjects were unable to continue holding their breath, they were allowed to stop immediately.
Facial videos were captured with a Logitech C920n HD PRO webcam at 30 frames per second (fps), and SpO2 data were acquired with an OxyTrue A pulse oximeter.
This study was approved by the NEC Ethical Review Committee for the Life Sciences (approval number: 21-017).

Since SpO2 data are acquired every four seconds in the pulse oximeter, we obtained one sample point per second by linear interpolation.
The spatio-temporal map consisted of 300 frames ($=T$) from 10 seconds of facial video, and the corresponding output was 10 seconds ($={D}_{\rm SpO2}$) of SpO2 data.
We performed a five-fold cross-validation, where 40 subjects were used for training and 10 subjects for test data.
Thus, we verified the estimation performance for new subjects not included in the training data, which is consistent with practical use scenarios.
In the training data, the spatio-temporal maps and corresponding SpO2 data were generated by overlapping in 2-second steps (60 frames). 
In the test data, the maps and SpO2 data were generated in 10-second steps (300 frames) without overlapping.
We used a random 20\% of the training data for validation data.
We trained for 50 epochs and used the model at the epoch with the lowest validation loss for testing.
During training, we used the Adam optimizer~\cite{kingma2014adam} with a learning rate of 1e-4.
In the training and validation data, the time-series data were normalized to mean 0 and variance 1 for each subject.
In the test data, for each subject, the current spatio-temporal maps were normalized using the mean and standard deviation of the current and prior time-series data.
SpO2 data were feature scaled to 0--1 within the range of possible SpO2 values (85--100\%).

\vspace{-5pt}
\subsection{Comparative Methods}
To verify the effectiveness of the proposed method, we compared it with RoR~\cite{rahman2019non,casalino2020mhealth}, linear regression (LR)~\cite{tian2022multi}, support vector regression (SVR)~\cite{tian2022multi}, MultiPhys~\cite{akamatsu2022heart}, CNN w/o pretrain, CNN, and CNN-Early.
In RoR, we used a 10-second window size, and $A$ and $B$ in Eq. (3) were set to the best parameters by using the training data.
In LR and SVR, we used the same explanatory variables as in~\cite{tian2022multi} for the DC and AC components extracted by our filtering processes.
Since a single ROI is used in RoR, LR, and SVR, we utilized an ROI that includes the nose and cheeks, which are used in~\cite{rahman2019non,casalino2020mhealth}.
MultiPhys is a CNN-based multimodal variational autoencoder that can estimate heart rate and SpO2 from facial videos.
We trained MultiPhys by additionally using heart rate measured from the pulse oximeter and transformed the shape of our spatio-temporal map to fit its network.
MultiPhys optionally utilizes datasets containing a part of physiological parameters or facial videos of new subjects, but we do not use them to make fair comparisons.
CNN w/o pretrain and CNN are models in which spatio-temporal maps are input without considering the DC and AC components.
CNN w/o pretrain was trained from scratch without pre-training, which is similar to the approach in~\cite{DBLP:journals/corr/abs-2107-05087}.
CNN-Early is a model in which the DC and AC components extracted by our filtering processes are input as six channels via early fusion, which is similar to the approach used in previous contact-based SpO2 estimation method~\cite{ding2018measuring}.
While previous CNN-based methods~\cite{DBLP:journals/corr/abs-2107-05087,ding2018measuring} were designed to take a single ROI as input, we used CNN models that handle multiple ROIs for our comparative methods.
CNN w/o pretrain, CNN, and CNN-Early were trained with the loss function in Eq. (5).
In the proposed method, we evaluated the filter-based model (CNN-Filter) and end-to-end model (CNN-EndToEnd), as shown in Fig.~\ref{fig:overview}.
ResNet18 was used for all CNN models, and all methods except for CNN w/o pretrain were trained from a model pre-trained on ImageNet.
As evaluation metrics for SpO2 estimation, we used mean absolute error (MAE), root mean squared error (RMSE), and Pearson correlation coefficient (CorrCoef) between predicted and ground truth data, as in~\cite{DBLP:journals/corr/abs-2107-05087}.

\begin{table}[t]
\caption{Mean SpO2 estimation performance across 50 subjects.}
\vspace{5pt}
\label{table}
\begin{center}
\scalebox{0.95}{
\begin{tabular}{l|ccc}  \hline
Method & MAE & RMSE & CorrCoef  \\ \hline \hline
RoR~\cite{rahman2019non,casalino2020mhealth} & 1.37 & 1.66 & 0.173  \\
LR~\cite{tian2022multi} & 1.25 & 1.51 & 0.193\\
SVR~\cite{tian2022multi} & 1.25 & 1.50 & 0.143 \\
MultiPhys~\cite{akamatsu2022heart} & 1.30 & 1.58 & 0.145 \\
CNN w/o pretrain & 1.25 & 1.49 & 0.399  \\
CNN & 1.23 & 1.47 & 0.455 \\ 
CNN-Early & 1.18 & 1.42 & 0.451  \\ \hline
{\bf CNN-Filter} & {\bf 1.17} & {\bf 1.40} & 0.474 \\ 
{\bf CNN-EndToEnd} & 1.18 & {\bf 1.40} & {\bf 0.496}\\
\hline
\end{tabular}
}
\end{center}
\vspace{-10pt}
\end{table}

\vspace{-5pt}
\subsection{Results}
Table~\ref{table} lists the SpO2 estimation performances of the proposed and comparative methods.
Note that the hyperparameter $\alpha$ in CNN-EndToEnd is set to 0.1.
From the table, it is clear that the proposed methods (CNN-Filter and CNN-EndToEnd) outperform all comparative methods.
In the following, we evaluate the effectiveness of each component in the proposed methods.
First, by comparing CNN w/o pretrain with RoR, LR, SVR, and MultiPhys, we can see that our proposed CNN-based method is superior to previous SpO2 estimation methods.
These results suggest that the deep neural network improves the performance of the signal processing-based method (RoR) and regression models (LR and SVR).
In addition, the previous study~\cite{akamatsu2022heart} reported that MultiPhys tends to perform poorly for new subjects not included in the training data, and our results also show low SpO2 estimation performance for new subjects.
Second, the comparison between CNN and CNN w/o pretrain shows the effectiveness of ImageNet pre-training.
This may be because ImageNet pre-training improves generalization ability despite the domain gap between image classification and SpO2 estimation (this effectiveness has also been confirmed in heart rate estimation~\cite{niu2018synrhythm}).
Third, by comparing CNN-Early, CNN-Filter, and CNN-EndToEnd with CNN, we confirm the effectiveness of considering the DC and AC components.
By designing CNN models that follow the principle of SpO2 estimation, we can further improve the estimation performance of standard CNN.
Finally, the comparison between CNN-Filter and CNN-Early shows that it is better to build two CNN models for the DC and AC components and fuse them than to input these components into a single model.
CNN-Filter and CNN-EndToEnd have comparable estimation performance, but CNN-EndToEnd is better in terms of correlation coefficient.

Figure~\ref{fig:spo2} presents examples of SpO2 estimation results using CNN-EndToEnd.
In subjects 1, 5, 19, 24, 45, and 47, SpO2 drops in accordance with the three breath holds, and the proposed method is able to capture this trend of the SpO2 data.
On the other hand, in subjects 38 and 44, the ground truth of SpO2 data sometimes does not drop during the three breath holds.
Even in such cases, the proposed method can successfully estimate the trend of the SpO2 data.
Although SpO2 estimation for new subjects not included in the training data is a difficult task~\cite{DBLP:journals/corr/abs-2107-05087,tian2022multi}, we find that our estimation results are relatively good even for new subjects.

\begin{figure}[t]
\begin{center}
\includegraphics[scale=0.5]{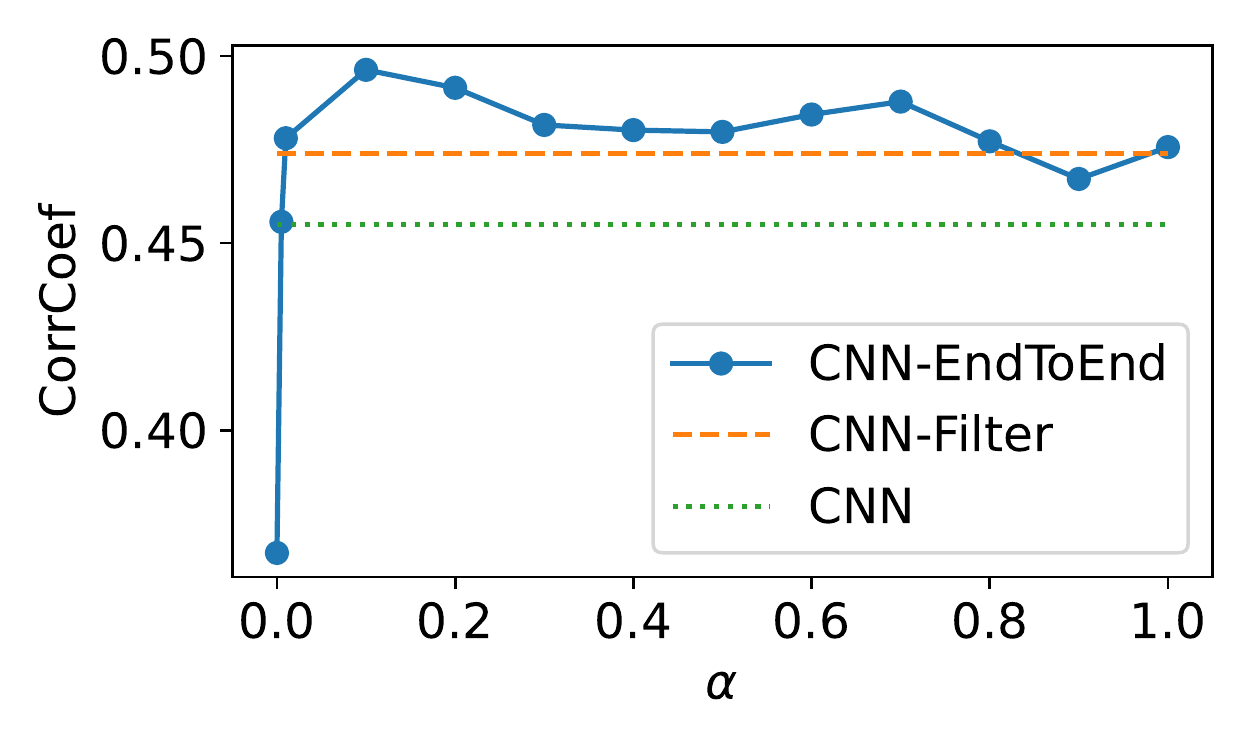}

\end{center}
\vspace{-25pt}
\caption{Relationship between hyperparameter $\alpha$ and SpO2 estimation performance in CNN-EndToEnd, CNN-Filter, and CNN.}
\label{fig:alpha}
\end{figure}

We also evaluate the estimation performance with respect to hyperparameter $\alpha$ in CNN-EndToEnd.
$\alpha$ controls the strength of the SpO2 estimation and the DC and AC components estimation in the loss function in Eq. (6).
When $\alpha$ is 0, only the loss of SpO2 estimation is calculated without considering DC and AC components estimation.
As $\alpha$ increases, the strength of the DC and AC component estimation becomes larger.
Figure~\ref{fig:alpha} shows the SpO2 estimation performance (correlation coefficient) when changing $\alpha$.
Since the estimation performance is lowest when $\alpha$ is 0, we can see the importance of considering DC and AC components estimation.
When $\alpha$ is set too large, the strength of the SpO2 estimation term in Eq.~(6) is reduced, resulting in poor estimation performance.
By tuning $\alpha$ to around 0.1, CNN-EndToEnd can achieve a higher estimation performance than CNN-Filter and CNN.
This result demonstrates the effectiveness of the end-to-end model that simultaneously trains SpO2 estimation and DC and AC components estimation.
Furthermore, CNN-EndToEnd can reduce the computational cost of the filtering processes in CNN-Filter during inference.

\section{Conclusion}
This paper presented a method for estimating SpO2 from facial videos via the DC and AC components of spatio-temporal maps.
We proposed filter-based and end-to-end CNN models that consider the DC and AC components of three-channel time-series data.
Experimental results with data from 50 subjects showed that the proposed method outperforms state-of-the-art SpO2 estimation methods.

\ninept
\bibliographystyle{IEEEbib}
\bibliography{main}
\end{document}